\documentstyle[12pt]{article}
\newcommand{\newc}{\newcommand}
\newc{\beq}    {\begin{equation}}
\newc{\eeq}    {\end{equation}}
\newc{\beqa}    {\begin{eqnarray}}
\newc{\eeqa}    {\end{eqnarray}}
\newc{\bibi}    {\bibitem}

\begin{document}

\begin{titlepage}
\title{
 Oscillating Inflation with a non-minimally coupled scalar field  }

\author{ Jae-weon Lee,\\
 Seoktae Koh,\\
 Chanyong Park,\\
 Sang Jin Sin, \\
 and\\
 Chul H. Lee,\\
      \it Department of Physics,
      \\ \it          Hanyang University,
                Seoul, 133-791, Korea
        \\
\\
}

\date{}
\setcounter{page}{1}
\maketitle
\vspace{-15cm}
\vspace{14cm}

The oscillating inflation model recently proposed by
Damour and Mukhanov is investigated
with a non-minimal coupling.
  Numerical study confirms
an inflationary behavior and  the density
perturbation is obtained.
A successful inflation requires
the gravity-dilaton coupling to be small.

PACS numbers: 98.80.Cq
\maketitle

\end{titlepage}

\newpage

 Recently, Damour and Mukhanov \cite{damour}  proposed
an inflation model with oscillating inflaton fields
in the Einstein gravity  context.
Meanwhile, dilaton-like fields are  natural candidates for
inflaton fields, because generally inflatons are supposed to be
gauge singlets. Furthermore, the extended gravity sector
 is common to   unified theories
such as supergravity, superstring/M-theory
 and Kaluza-Klein theory\cite{sugra}.
 These dilaton fields might be stabilized
 by some potential, which could be an inflaton potential.
In these respects, in this paper, we study an oscillating
inflation with non-minimal coupling.

 We consider an  action of non-minimally coupled scalar theory
which is given by\cite{acceta}
\beq
 S=\int \sqrt{-g} d^4x \left [ -U(\phi)R
+\frac{1}{2}\partial_\mu\phi
\partial^\mu \phi -V(\phi) \right ],
\label{action}
 \eeq
 where $\phi$ is a dilaton-like inflaton field
 and $U(\phi)=M_P^2/16\pi+\xi
\phi^2/2$.  We adopt the following inflaton potential
 \beq
V(\phi) = \frac{A}{q}\left [ \left
(\frac{\phi^2}{\phi_0^2}+1\right )^{q/2}-1\right]
\label{V}
 \eeq
 which was suggested by Damour and Mukhanov\cite{damour}.
Here $A$,$q$ and $\phi_0$ are  constants.
The equations of motion for the scale factor $a$ and $\phi$
 from the action (Eq.(\ref{action}))
with the metric $ds^2=-dt^2 +a(t)^2 dx^2$ are
\beq
3H^2\left[ \frac{M_P^2}{8\pi }+\xi \phi^2 \right ]=
\frac{\phi'^2}{2}+V(\phi)-6\xi H\phi \phi '
\label{h2}
\eeq
and
\beq
{\phi}{''}+3 H{\phi}'-6\xi (H'+2H^2)\phi+\frac{dV}{d\phi}=0,
\label{p2}
\eeq
where the prime denotes $d/dt$ and $H\equiv a'/a$.
To see whether oscillating inflation really happens in this
model,
 we  do numerical study.
A typical signal of an oscillating inflation is an increasing
and wiggling curve of $aH$ versus $t$.
The result of numerical calculation shows this curve(Fig.1.)
and confirms inflationary behaviours.

Now let us
 show that  $\xi<q\ll1$ is a good parameter range
for our model with the Damour-Mukhanov potential
 and calculate the density perturbation.
 The COBE observation  requires $\delta_{
H} \simeq 2 \times 10^{-5}$.
Since the oscillating inflation phase is generally short,
the observed perturbation is supposed to be generated during
a slow-roll phase.
From Eq.(\ref{h2}) and Eq.(\ref{p2}) and with
slow-roll conditions one can obtain\cite{slow}
\beq
3 H \phi' \simeq \frac{1}{1+ \xi \kappa^2 \phi^2(1+6\xi)}
\left [ 4 \xi \kappa^2  \phi V(\phi)- (1+
\xi \kappa^2 \phi^2)\frac{dV}{d\phi}
\right ],
\label{slowroll}
\eeq
where $\kappa^2=8\pi /M_P^2$.
Since during  the slow-roll phase  $\phi\gg \phi_0$
, one can approximate the Damour-Mukhanov potential as
\beq
V(\phi)\simeq \frac{A}{q} (\frac{\phi}{\phi_0})^q.
\eeq
However, for $q\ll 1$ which is required for the oscillating
inflation\cite{damour}, such a polynomial potential
with small exponent brings the following
problem with non-minimal coupling\cite{maeda};
When $q\ll 1$ the first term in Eq.(\ref{slowroll}) is larger than the second one
 and the former gives
a force which prevents $\phi$ rolling down to
the potential minimum.  One can overcome this problem
by simply choosing the parameters  satisfying
\beq
\xi \kappa^2
\phi^2 < q/(4-q) \ll1.
\label{condition}
\eeq
Since  for the  potential
the number of e-folds of expansion during the slow-roll is given by

\beq
N\simeq -\frac{8\pi}{M_P^2}\int \frac{V}{dV/d\phi}d\phi \simeq
 \frac{4\pi \phi^2}{q M_P^2},
\eeq
 one can obtain a sufficient expansion
with initial $\phi=\phi_N \sim M_P$.
Then the above condition(Eq.(\ref{condition})) becomes $\xi < 1/8N$.
Furthermore, it is known that for $\xi \ll 1$,  the density perturbation
is proportional to $H^2/|\phi'|$  as usual\cite{density}.
Therefore, in this limit, \cite{liddle}
\begin{equation}
\delta_{ H}^2 \simeq \frac{512\pi}{75} \, \frac{A}{q^3 {M_P}^6}
        \, \frac{\phi^{2+q}}{\phi_{0}^q}
\left [ 1+\xi (3 q -\frac{3\kappa^2\phi^2}{2}+\frac{\kappa^2 \phi^2}{q})
\right ] ^2
\end{equation}
is a good approximation up to $O(\xi)$
 within the mentioned parameter
ranges($\xi<q\ll1$).
The term with $\xi$ represents
the effect of non-minimal coupling which is dominated by
the last term, $\xi \kappa^2\phi^2/q$ for $q\ll 1$.
Therefore in the marginal case( $\xi\kappa^2\phi^2/q \simeq \frac{1}{4}$)
this could contribute about $\frac{1}{4}$ to  the factor of $\delta_H$.
Let us approximate $\delta_H^2$ as
\beq
\delta_H^2(\xi \neq 0) \simeq \delta^2_H(\xi=0)\left [ 1+\frac{2\xi \kappa^2\phi^2}{q}
\right ].
\eeq
Since one can easily change the magnitude of $\delta_H$ by
choosing constants in the potential such as $A $ and $\phi_0$,
one need to observe the spectral index $n$ of the perturbation
rather than the magnitude of it.
The background radiation observation
 MAP\cite {map} and Planck\cite{planck} will be a discriminator between
many inflation  models.
Since
\beq
 n=1+\frac{25}{4}\frac{d \delta_H^2}{dlnk},
\eeq
where $k$ is a wave number of the perturbation \cite{lyth},
the deviation of $n$ between the oscillating inflation with  Einstein gravity
and with the non-minimal coupling is
\beqa
\Delta n &=&\frac{25}{4} \frac{d[ \delta_H^2(\xi\neq 0)- \delta_H^2(\xi=0) ]}
{dlnk} \\ \nonumber
&\simeq& \frac{25}{2}\frac{d\phi}{dlnk}\frac{d(\xi\kappa^2\phi^2/q)}
{d\phi} \delta_H^2(\xi=0)+\frac{2\xi \kappa^2\phi^2}{q}\left [ n(\xi=0)-1 \right ],
\eeqa
where $d\phi/dlnk=-(M_P^2/8\pi V)dV/d\phi$.
The first term contributes $ -10^{-10} \xi$ to $\Delta n$
which is too small to be observed by the satellites.
Using $1-n \simeq (q+2)/2N$ for chaotic inflation
with $V\sim \phi^q$ \cite{n}, one can find
that even in the marginal case
the second term is about $-\xi \kappa^2\phi^2/30q\simeq 1/120$,
which is  rather smaller than the accuracy of Planck $\Delta n \sim
0.02$. So it is hard to expect the observations could
distinguish the  two models near future.

Now let us check inflation conditions during the  oscillating phase.
The scale factor also satisfies the following equation;
\beq
\frac{a''}{a}=\frac{\kappa^2}{1+\xi \kappa^2  \phi^2}
\left [ -\phi'^2(1+3\xi) +V(\phi) -3 \xi \phi( H\phi'+\phi'')
\right ].
\label{a2}
\eeq
We adopt a reasonable assumption that during the oscillation
the friction term is negligible ( $\phi'' \gg 3H\phi'$) and
the effective mass squared of
the field (See Eq.(\ref{p2}))
\beq
m^2\equiv \frac{d^2 V(\phi)}{d\phi^2}-6\xi(H'+2H^2) \gg H^2.
\label{m}
\eeq
Since the typical period of an oscillation is $1/m$,
one can  ignore the variation of $H$
 ( and the $H'$ dependent term ) during a single oscillation.
One can also easily find that in Eq.(\ref{m}), $\xi H^2$ dependent term
should be small relative to $d^2V/d\phi^2$, especially when $\xi \ll 1$.
So it  is a good approximation that $\phi'' \simeq -dV/d\phi$.
Since we expect that the present value of $U(\phi)$ is equal to
$M_P/16\pi$, the present value of  $\phi$ should be $0$.
So the limit
$ \xi \kappa^2 \phi^2 \rightarrow 0$
  is a reasonable assumption near the potential
minimum  and for $\xi \ll 1.$
In this limit and  with $\langle \phi \phi'' \rangle=
\langle -\phi'^2 \rangle $ the inflation condition is
\beqa
 \left \langle \frac{a''}{a} \right \rangle
& \simeq &
\left \langle {\kappa^2}
\left [ -\phi'^2(1+3\xi) +V(\phi) -3 \xi \phi( H\phi'+\phi'')
\right ] \right \rangle  \nonumber        \\
& \simeq &
\kappa^2 \left \langle - \phi'^2 +V(\phi) \right  \rangle > 0,
\eeqa
   where the bracket denotes a time average during an oscillation
period.

With $\phi''= -dV/d\phi$ this reduces to
\beq
\left \langle{V} - \phi  \frac{dV}{d\phi}
 \right \rangle > 0 ,
\eeq
which is  the result of Ref.\cite{damour}.
Of course, this is simply due to the fact  that
  in this limit the system becomes
that with  Einstein gravity.
So, our model is self-consistent  at least when $\xi <q\ll 1$.

In summary, in the context of the non-minimal coupling, we
investigate
 the oscillating inflation and calculate the density perturbation
during the slow-roll phase with the Damour-Mukhanov potential.

\vskip 1cm

This work was supported in part by KOSEF and Korea research
foundation(BSRI-98-2441).

\newpage

\section*{ Figure Caption }
Fig.1 \\
 $aH$ vs. $t$ from numerical solutions of the field
equations with $\xi=0.001$,$q=0.1$,$\phi_0=0.01$ and $A=10^{-14}$.
All the quantities are denoted in the natural units where $M_P=1$.

\newpage
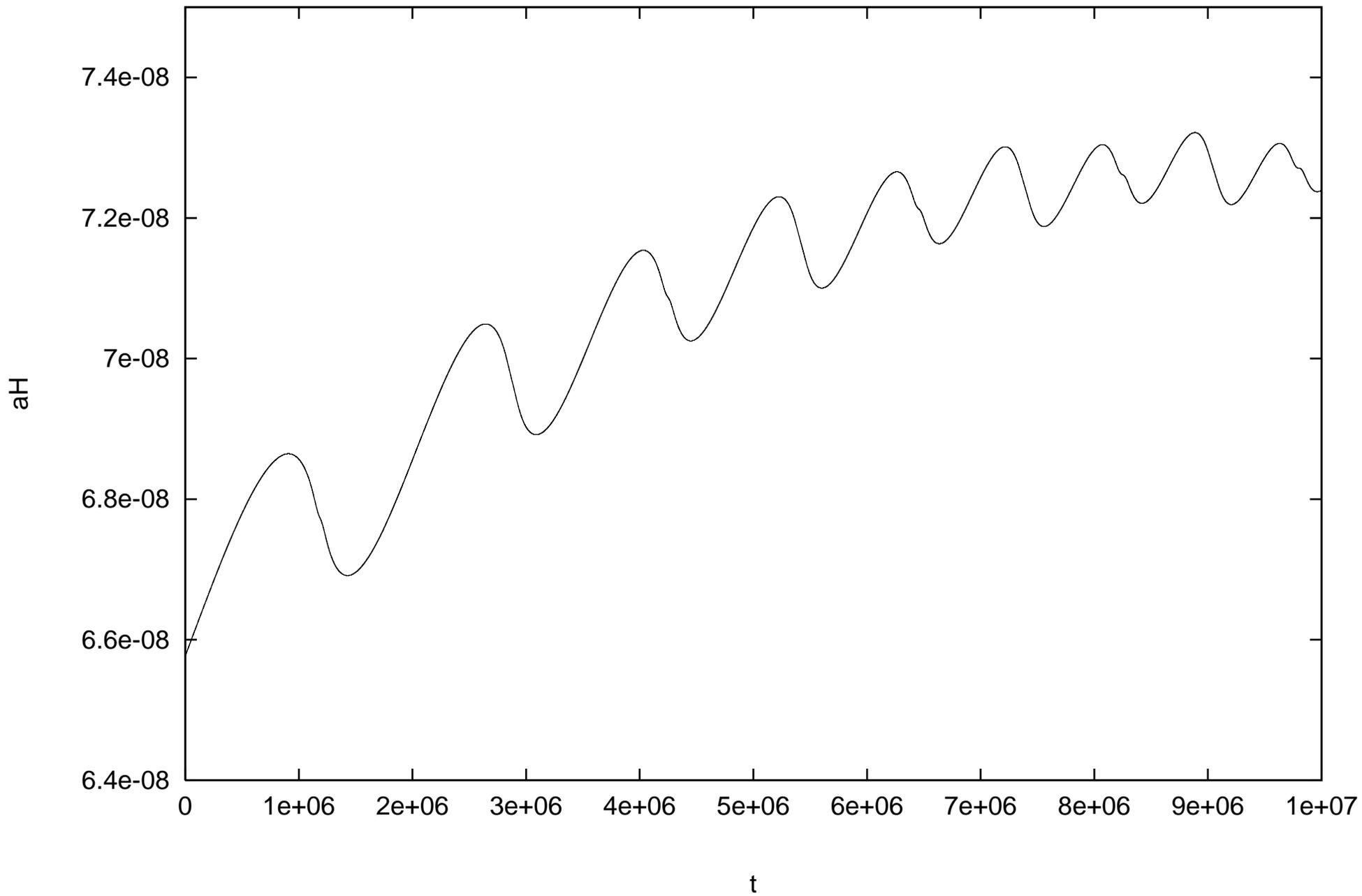
\end{document}